\documentstyle[preprint,aps,epsfig]{revtex}
                   
\begin{document}
\draft
\title {Magneto-controlled nonlinear optical materials  }
\date {\today}

\author {J. P. Huang\footnote{Electronic address: jphuang@mpip-mainz.mpg.de}}
\address {Department of Physics, The Chinese University of Hong Kong, 
 Shatin, New Territories, Hong Kong  and Max Planck Institute for Polymer Research, Ackermannweg 10, 
 55128 Mainz, Germany}
\author {K. W. Yu}
\address {Department of Physics, The Chinese University of Hong Kong, 
 Shatin, New Territories, Hong Kong}


\maketitle

\begin{abstract}

We exploit theoretically a magneto-controlled nonlinear optical
material which contains ferromagnetic nanoparticles with a non-magnetic
metallic nonlinear shell in a host fluid. Such an optical material can
have anisotropic linear and nonlinear optical properties and a giant
enhancement of nonlinearity, as well as an attractive figure of merit.




\vskip 5mm
\end{abstract}

\newpage


Finding nonlinear optical materials with large nonlinear susceptibilities and fast responses is up to now a challenge~\cite{DodenbergerN92,FischerPRL95,MishinaNL01,HuangAPL04}. Many applications of nonlinear optics that have been demonstrated under controlled laboratory conditions could become practical for technological uses if such materials were available.  The most common way to obtain a nonlinear optical material is to search for materials in which the components possess an inherently large nonlinear response~\cite{DodenbergerN92}. In contrast, in this Letter we shall  exploit theoretically a  nonlinear optical material whose nonlinear optical properties and nonlinearity enhancement can be tuned by applying an external magnetic field - thus called {\it magneto-controlled nonlinear optical materials}. Devices that could benefit from these  materials include optical switches, optical limiters,  etc.
Ferromagnetic nanoparticles, typically consisting of magnetite or cobalt, have a typical diameter of 10 nm, and carry a permanent magnetic moment (e.\,g., of the strength $\sim 2.4\times 10^4\,$$\mu_B$ for magnetite nanoparticles, where $\mu_B$ denotes the Bohr magneton)~\cite{Odenbach02}.
As the ferromagnetic nanoparticles are suspended in a host fluid like water, they can easily form particle chains under the application of external  magnetic fields~\cite{Odenbach02}, thus yielding a magnetic-field-induced anisotropical structure. 
Recently, a non-magnetic golden shell was used to enhance the stability of the ferromagnetic nanoparticle against air and moisture~\cite{BonnemannICA03}. In this work, we shall show that the effective nonlinear optical response of the suspension which contains ferromagnetic nanoparticles with  metallic nonlinear shells can be enhanced significantly due to the effect of the magnetic-field-induced anisotropy. 
 For the research on nonlinear optical responses, the introduction of a controllable element (e.g., external magnetic field) should be expected to open a fascinating field of new phenomena.


The third-order nonlinear susceptibility $\chi_s$ of metallic (say typically, noble metals like gold and silver) shells  is very large when compared to that of the  magnetite or cobalt core and the host  fluid like water. 
Let us start by considering ferromagnetic linear nanoparticles of linear dielectric constant $\epsilon_1''$ coated with a non-magnetic metallic nonlinear shell of $\epsilon_1'$ and $\chi_s$ which are suspended in a linear host fluid of $\epsilon_2$. That is, in the shells, there is a nonlinear relation between the displacement ${\bf D}_s$ and the electric field ${\bf E}_s$,
${\bf D}_s = \epsilon_1'{\bf E}_s+\chi_s|{\bf E}_s|^2{\bf E}_s,$
where $\epsilon_1'$ is given by the Drude form,
$\epsilon_1'=1-\omega_p^2/[\omega(\omega+\gamma i)],$
where $\omega_p$ and $\gamma$ stand for the plasmon frequency and the relaxation rate, respectively, and $\omega$ denotes the frequency of the incident light.
In what follows, the thickness of the shell and the radius of the core are respectively denoted as $d$ and $R$.
Next, we  restrict our discussion to the quasi-static approximation, under which the structured particle or the whole suspension can be regarded as an effective homogeneous one.
It is known that the effective third-order nonlinear susceptibility $\bar{\chi}$ of an area [in this work, the area represents the structured particle and the whole suspension, respectively, see Eqs.~(\ref{chi_1})~and~(\ref{chi_e}) below] is defined as~\cite{StroudPRB88,YuPRB93}
\begin{eqnarray}
\bar{\chi} = \frac{1}{V|E_0|^2E_0^2}\int_V\chi({\bf r})|\nabla\phi_0({\bf r})|^2[\nabla\phi_0({\bf r})]^2{\rm d}{\bf r},\label{def}
\end{eqnarray}
which is in terms of zeroth-order potential $\phi_0({\bf r})$ only, see Eqs.~(\ref{sol1}-\ref{sol2}) below. In Eq.~(\ref{def}) $E_0$ denotes the external applied electric field, $V$ the volume of the area under consideration, ${\bf r}$ the local position inside the medium ($r$ the distance from the
particle center to the point of interest),  and $\chi({\bf r})$  an ${\bf r}$-dependent third-order nonlinear susceptibility.
To obtain the effective nonlinear susceptibility of the structured particle which contains a linear core with a nonlinear shell, we should obtain the zeroth-order potentials which are actually obtained for the system in which the nonlinear characteristic of shells disappears,  $\chi_s=0$. 
Under the quasi-static approximation, the Maxwell equations read
\begin{equation}
\nabla\times {\bf E} = 0\,\, {\rm and}\,\,
\nabla\cdot {\bf D} = 0,\label{M1}
\end{equation}
and hence ${\bf E} = -\nabla\phi$, where $\phi$ is an electric potential.
Solving Eqs.~(\ref{M1}) [or the corresponding Laplace equation $\nabla^2\phi=0$], we obtain the zeroth-order potentials for the core $\phi_0{}^c$, the shell $\phi_0{}^s$, and the host $\phi_0{}^h$
\begin{eqnarray}
\phi_0{}^c &=& -c_1E_0r\cos\theta, r<R,\label{sol1}\\
\phi_0{}^s &=& -E_0(c_2r-c_3r^{-2})\cos\theta, R<r<R+d, \\
\phi_0{}^h &=& -E_0(r-c_4r^{-2})\cos\theta, r>R+d,\label{sol2}
\end{eqnarray}
where $\theta$ is the angle between the external field and the line joining the particle center and the point under investigation, and the coefficients $c_1$, $c_2$, $c_3$, and $c_4$ are determined by the appropriate boundary conditions. 
Owing to Eq.~(\ref{def}), {\it the effective third-order  nonlinear susceptibility  of the structured particle} $\chi_1$  can be given by
\begin{equation}
\chi_1\frac{\langle|\nabla\phi_0({\bf r})|^2[\nabla\phi_0({\bf r})]^2 \rangle_{r\le R+d}}{|{\bf E}_0|^2{\bf E}_0^2} = f\chi_s\frac{\langle|\nabla\phi_0({\bf r})|^2[\nabla\phi_0({\bf r})]^2 \rangle_{R<r\le R+d}}{|{\bf E}_0|^2{\bf E}_0^2},
\end{equation}
where $f$ is the volume ratio of the shell to the core. Thus, we obtain
\begin{equation}
\chi_1 = \chi_s\frac{\beta}{\beta'},\label{chi_1}
\end{equation}
where
$\beta = (3/5)[1/(1-f)^{1/3}-1]|z|^2z^2(5+18x^2+18|x|^2+4x^3+12x|x|^2+24|x|^2x^2)$
and
\begin{equation}
\beta' =\left |\frac{\epsilon_2}{\epsilon_2+(\alpha/3)(\epsilon_1-\epsilon_2)}\right |^2\left ( \frac{\epsilon_2}{\epsilon_2+(\alpha/3)(\epsilon_1-\epsilon_2)}\right )^2 \label{beta1}
\end{equation}
with
$
x=(\epsilon_1''-\epsilon_1')/(\epsilon_1''+2\epsilon_1')
$
and
$
z=(1/3)[\epsilon_2(\epsilon_1''+2\epsilon_1')]/\{\epsilon_1'[\epsilon_2+(\alpha/3)(\epsilon_1''-\epsilon_2)]\}.
$
In Eq.~(\ref{beta1}), the effective linear dielectric constant $\epsilon_1$   of each structured particle can be determined by the well-known Maxwell-Garnett formula with a high degree of accuracy
\begin{equation}
\frac{\epsilon_1-\epsilon_1'}{\epsilon_1+2\epsilon_1'}=(1-f)\frac{\epsilon_1''-\epsilon_1'}{\epsilon_1''+2\epsilon_1'}.\label{particle}
\end{equation}
It is worth noting that for the above derivation a local field factor $\alpha$ has been introduced, see Eq.~(\ref{beta1}). In detail, $\alpha$ denotes the local field factors $\alpha_L$ and $\alpha_T$ for longitudinal and transverse field cases, respectively. Here the longitudinal (or transverse) field case corresponds to the fact that the $E-$field of the light is parallel (or perpendicular) to the particle chain.   Similar factors in electrorheological fluids were measured by using computer simulations~\cite{Martin98,Martin98-2}, and obtained theoretically~\cite{LoPRE01,HuangPRE04} according to the Ewald-Kornfeld formulation. 
There is a sum rule for  $\alpha_L$ and $\alpha_T$,  $\alpha_L+2\alpha_T=3$~\cite{LandauECM84}.  The parameter $\alpha$ measures the degree of 
anisotropy, which is induced by the applied magnetic field $H$. More precisely, the degree of the field-induced anisotropy is measured by how much $\alpha$ deviates from unity, $1<\alpha_T<3$ for transverse field cases and  $0<\alpha_L<1$ for longitudinal field cases. As $H$ increases $\alpha_T$ and $\alpha_L$ should tend to $3$ and $0$, respectively, which is indicative of the formation of more and more particle chains
as evident in experiments~\cite{Odenbach02}.
So, a crude
estimate of $\alpha$ can be obtained from the contribution of chains~\cite{Rasa99}, namely, 
$\alpha = [4\pi (d+R)^3/p]\sum_{n=1}^{\infty}n\gamma_n(H)g_n,$
where $p$ denotes the volume fraction of the structured particles in the suspension,  $g_n$  the depolarization factor for a chain with $n$ structured particles, and $\gamma_n(H)$  the density of the chain which is a 
function of $H$. It is noteworthy that for given $p$ $\gamma_n(H)$ also depends on the dipolar coupling constant which relates the dipole-dipole interaction 
energy of two contacting particles to the thermal energy.  
Now, the system of interest can be equivalent to the one in which all the particles with linear dielectric constant $\epsilon_1$ [Eq.~(\ref{particle})] and nonlinear susceptibility $\chi_1$ [Eq.~(\ref{chi_1})] are embedded in a host fluid with $\epsilon_2$. For the equivalent system, it is easy to solve the corresponding Maxwell equations [Eqs.~(\ref{M1})], in order to get the zeroth-order potentials in the particles and the host.   According to Eq.~(\ref{def}), we obtain {\it the effective third-order  nonlinear susceptibility of the whole suspension} $\chi_e$ as
$\chi_e = p\chi_1\beta',$
 which can be rewritten as
\begin{equation}
\chi_e =  p\chi_s\beta,\label{chi_e}
\end{equation}
The substitution of $\alpha=1.0$ (i.e., the isotropic limit) into  Eq.~(\ref{chi_e}) yields  the same expression as derived in Ref.~\cite{YuPRB93} in which the dielectric constants of the core and shell of structured particles were, however, assumed to be real rather than complex.
On the other hand, the effective linear dielectric constant of the whole suspension under present consideration $\epsilon_e$ can be given by the developed Maxwell-Garnett approximation which works for suspensions with field-induced anisotropic structures~\cite{LoPRE01}
\begin{equation}
\frac{\epsilon_e-\epsilon_2}{\alpha\epsilon_e+(3-\alpha)\epsilon_2}=p\frac{\epsilon_1-\epsilon_2}{\epsilon_1+2\epsilon_2}.\label{aniso}
\end{equation}

For numerical calculations, without loss of generality we take $f=0.65$,  $p=0.2$,   $\epsilon_1''=-25+4i$,  $\epsilon_2=1.77$ (dielectric constant of water), and $\gamma = 0.01\omega_p.$ We further see $\chi_s$ to be a real and positive frequency-independent constant, in order to focus on the nonlinearity enhancement.
Figures~1~and~2 display the linear optical absorption ${\rm Im}(\epsilon_e)$, the enhancement of the
third-order optical nonlinearity   $|\chi_e|/\chi_s$, and the figure of merit (FOM) $|\chi_e|/[\chi_s{\rm Im}(\epsilon_e) ]$, as a function of normalized frequency $\omega/\omega_p$, for (Fig.~1) longitudinal and (Fig.~2) transverse  field cases. Here the frequency $\omega$ is normalized by $\omega_p$ (rather than a specific value of $\omega_p$), so that the result could be valid for general cases. As mentioned before, $\alpha = 1.0$ corresponds to the isotropic limit. In this case, there is no external magnetic field, and hence all the structured particles are randomly suspended. The figures show that the existence of nonlinear shells causes an enhancement of nonlinearity to appear, see Fig.~1(b) and Fig.~2(b), thus yielding a large FOM, see Fig.~1(c) and Fig.~2(c). Such a nonlinearity enhancement induced by shell effects was already reported~\cite{YuPRB93}. 
The main feature of Figures~1~and~2 is the effects of external magnetic fields. As $\alpha_L$ changes from $1.0$, to $0.6$, and to $0.2$, (namely, as $\alpha_T$ varies from $1.0$, to $1.2$, and to $1.4$) the external magnetic field is adjusted from zero, to low strength, and to high strength. Due to the interaction between the ferromagnetic nanoparticles and the magnetic field, more and more particle chains are caused to appear naturally, thus yielding a magnetic-field-induced anisotropic structure in the suspension. It is evident to observe that the plasmon peak is caused to be blue-shifted for longitudinal field cases as the magnetic field increases. However, for transverse field cases, the plasmon peak displays a red-shift for the increasing  magnetic field. In other words, the optical absorption is induced to be anisotropic due to the application of the external magnetic field which produces an anisotropic structure. In fact, the optical absorption arises from the surface plasmon resonance, which is
obtained from the imaginary part of the effective dielectric constant.
For single metallic particles in the dilute limit, it is well known that
there is a large absorption when the resonant condition $\epsilon_1'
+ 2\epsilon_2 = 0$ is fulfilled. When there is a larger volume fraction $p$ of
structured particles and an anisotropy $\alpha$ of the suspension, the effective dielectric constant
should be obtained from Eq.~(\ref{aniso}),
thus yielding a modified resonant condition 
$(1-p\alpha)\epsilon_1 + (2+p\alpha)\epsilon_2 = 0.$ So, the
resonant frequency becomes larger (smaller) than the isotropic limit
($\alpha=1$) when $\alpha$ becomes smaller (larger) than $1$. In other words,
there is a blue (red) shift for the longitudinal (transversal) field cases.
More interestingly, for longitudinal field cases, a giant enhancement of nonlinearity is shown as the magnetic field increases, see Fig.~1(b). In detail, the nonlinearity enhancement of a high-field case (say, $\alpha = 0.2$) can be of five orders of magnitude larger than that of the zero-field case ($\alpha=1.0$). Inversely, a reduction of nonlinearity is found for transverse field cases, see Fig.~2(b). The magnitude of the nonlinearity reduction is very small in the transverse field case, when compared to that of the nonlinearity enhancement in the longitudinal field case. 
Owing to the giant enhancement of nonlinearity [see Fig.~1(b)], the FOM becomes much more attractive for longitudinal field cases [see Fig.~1(c)]. The FOM of a high-field case (say, $\alpha=0.2$) can even be ten-order-of-magnitude enhanced in the longitudinal field case. However, the effect of the magnetic field on the FOM for  transverse field cases seems to be uninteresting since the FOM is caused to be decreased slightly due to the nonlinearity reduction shown in Fig.~2(b). 
Since the permanent magnetic moment of the
magnetite nanoparticles $m$ is approximately $2.4\times 10^4\,$$\mu_B$~\cite{Odenbach02}, we can estimate the threshold magnetic field  $H_c= 14.3\,$kA/m (or threshold magnetic induction $B_c=0.018\,$T) above which the corresponding magnetic
energy can overcome the thermal energy $1/40\,$eV so as to obtain appreciable
anisotropy. 
Besides the magnetic energy, we should also compare the interaction energy. For instance, for two
touching magnetite nanoparticles, the interaction between them is proportional to $m^2/[2(d+R)]^3$,
assuming the two structured particles to be in a head-to-tail alignment.   Since the magnetic moment $m$ goes as $(2R)^3$,
 the interaction energy could vary as $[2R^2/(d+R)]^3$.
In order to break up the two touching nanoparticles, the thermal energy should be
larger than the interaction energy.
So, threshold field $H_c= 14.3\,$kA/m serves as an upper estimate. Nevertheless, for cobalt nanoparticles, the threshold field $H_c$ should be lower due to  larger  permanent magnetic moments.
To sum up, by including a metallic nonlinear shell in the system, one can tune the linear and nonlinear optical properties by applying a magnetic field. 
 Such a proposed magneto-controlled nonlinear  optical material can   serve as   optical materials which have anisotropic nonlinear optical properties and a giant enhancement of nonlinearity, as well as an attractive FOM.  






{\it Acknowledgments}.
We thank Prof. K.  Yakubo, Prof. T. Nakayama and Dr. C. Holm for fruitful discussions.
This work was supported by the Research Grants Council of the Hong Kong  SAR Government,  by  the DFG under Grant No. HO 1108/8-4 (J.P.H.), by  the Alexander von Humboldt Foundation in Germany (J.P.H.), and in part by the Grant-in-Aid for Scientific Research organized by Japan Society for the Promotion of Science.



\begin{figure}[h]
\caption{ (a) The linear optical absorption ${\rm Im}(\epsilon_e)$, (b) the enhancement of the
third-order optical nonlinearity   $|\chi_e|/\chi_s$, and (c)  the FOM   $|\chi_e|/[\chi_s{\rm Im}(\epsilon_e) ]$ versus the normalized incident angular frequency
$\omega/\omega_p$, for various strengths of the external magnetic field which are represented by  local-field factors $\alpha_L$, for longitudinal field cases (L).
  }
\end{figure}

\begin{figure}[h]
\caption{Same as Fig.~1, but for transverse field cases (T). }
\end{figure}

\newpage
\centerline{\epsfig{file=fig1.eps,width=200pt}}
\centerline{Fig.1./Huang and Yu}

\newpage
\centerline{\epsfig{file=fig2.eps,width=200pt}}
\centerline{Fig.2./Huang and Yu}

\end{document}